\documentclass[pra,twocolumn,superscriptaddress,10pt,noshowpacs]{revtex4}
\usepackage[english]{babel}
\usepackage[T1]{fontenc}
\usepackage[utf8]{inputenc}
\usepackage{graphicx,epstopdf}
\usepackage{amssymb}
\usepackage{amsmath}

\usepackage{amsfonts}
\usepackage{bbm}
\usepackage{color}
\usepackage{latexsym}
\usepackage{caption}
\usepackage{subcaption}
\usepackage{times,txfonts}

\newcommand{\beq}{\begin{equation}}
\newcommand{\eeq}{\end{equation}}
\newcommand{\bea}{\begin{eqnarray}}
\newcommand{\eea}{\end{eqnarray}}

\begin{document}
\title{Thermodynamics and Quasinormal Modes of the Dymnikova Black Hole in Higher Dimensions}

\author{M. H. Mac\^{e}do}
\email{matheus.macedo@fisica.ufc.br}
\affiliation{Universidade Federal do Cear\'a (UFC), Departamento de F\'isica,\\ Campus do Pici, Fortaleza - CE, C.P. 6030, 60455-760 - Brazil.}
\author{J. Furtado}
\email{job.furtado@ufca.edu.br}
\affiliation{Universidade Federal do Cariri (UFCA), Av. Tenente Raimundo Rocha, \\ Cidade Universit\'{a}ria, Juazeiro do Norte, Cear\'{a}, CEP 63048-080, Brasil}
\affiliation{Department of Physics, Faculty of Science, Gazi University, 06500 Ankara, Turkey}

\author{G. Alencar}
\email{geova@fisica.ufc.br}
\affiliation{Universidade Federal do Cear\'a (UFC), Departamento de F\'isica,\\ Campus do Pici, Fortaleza - CE, C.P. 6030, 60455-760 - Brazil.}

\author{R. R. Landim}
\email{renan@fisica.ufc.br}
\affiliation{Universidade Federal do Cear\'a (UFC), Departamento de F\'isica,\\ Campus do Pici, Fortaleza - CE, C.P. 6030, 60455-760 - Brazil.}

\date{\today}

\begin{abstract}

In this study, we investigate the thermodynamic properties and quasinormal modes of Dymnikova black holes within the context of higher dimensions in Einstein's general theory of relativity. We calculate the thermodynamic parameters, including the Hawking temperature and heat capacity, which allowed us to investigate the black hole's stability. Lastly the quasinormal modes with the WKB formula were calculated.

\end{abstract}

\maketitle

\section{Introduction}

The modern idea of a black hole, that is, a region of space from which nothing can escape, dates back to Finkelstein's publication \cite{Finkelstein:1958}, in which he asserts that the event horizon is a one-way membrane through which causal events can only go in one direction.  In 2019, the first direct image of a black hole was captured by the Event Horizon Telescope \cite{EventHorizonTelescope:2019dse}. This reaching marked a significant milestone in astrophysics, providing concrete visual evidence of the existence of black holes, which had long been theorized but never directly observed.

Bekenstein postulated that black holes possess an entropy correlated with the area of the event horizon and, moreover, that the energy is related to mass of the black hole \cite{Bekenstein:1972, Bekenstein:1973, Bekenstein:1974}. Hawking revealed that black holes emit thermal radiation, and this radiation's temperature has similarity with the black hole's surface gravity \cite{Hawking:1975}.

The essential singularity in Einstein's gravity can be preventable if the strong energy condition is broken in the vicinity of a black hole center. Black holes without essential singularities, but with coordinate singularities, are called regular black holes. The investigation of Regular Black Holes has its origins in the works of Sakharov and Gliner \cite{Sakharov:1966, Gliner:1966}, who proposed that essential singularities could be circumvented by substituting the vacuum with a vacuum-like medium possessing a de Sitter metric. Later, Bardeen \cite{bardeen} proposed a model of regular black hole by promoting the mass of the Schwarzschild black hole to a position-dependent function. Since then, several regular black holes were proposed in the literature in different contexts, such as non-local gravity \cite{Christiansen:2022ebo}, BTZ bounce \cite{Furtado:2022tnb}, f(R) gravity \cite{Rodrigues:2015}, among others.

A very interesting black hole's regular solution was proposed by Dymnikova \cite{Dymnikova:1992} in the usual four dimensions of general relativity. It was assumed a specific form for the stress-energy momentum tensor and derived the vacuum non-singular black hole metric, such solution reduces to de Sitter solution for small $r$, for large $r$ behave like  Schwarzschild solution. The metric has a Cauchy horizon and an external horizon, but both are removable and the solution is regular at $r = 0$. 
 The Dymnikova solution exhibits regularity throughout, as evidenced by the behavior of scalar invariants and the Kretschmann invariant. These invariants maintain good behavior in all regions, including at $r = 0$. Consequently, the black hole does not lead to a singularity.

There are notable distinctions between black holes that exhibit singularities and those considered regular \cite{Lan:2023cvz}. Some of these distinctions arise directly from the absence of singularity, while others are intrinsic to the model itself. Such discrepancies have repercussions that manifest in various ways in observation, including notable effects such as gravitational waves, shadows, thermodynamic properties, and more.

Since it is impossible to directly observe the interior of a black hole, our research must focus on the dynamic and thermodynamic events that occur beyond the event horizons. Specifically, we aim to understand the variations in these phenomena when singularities are present compared to when they are not.

In recent years, several studies on black holes and related objects in higher dimensions have been conducted, and such research has been spurred by the development of string theory. For this reason, it is pertinent to study the D-dimensional generalization of Dymnikova black holes,  made recently by Paul \cite{Bikash Chandra Paul:2023}, and discuss their thermodynamical properties. The Dymnikova black hole has received significant attention in the recent past, e.g., the stability  of the Dymnikova black hole was performed by Nashed and Dymnikova \cite{Nashed:2003, Dymnikova:2005}, the study of quasinormal modes in Dymnikova black holes  were discussed in \cite{Konoplya:2023} and thermodynamic quantities was calculated in \cite{Sharif:2022}. More recently the Dymnikova-like regularization was studied in the context of traversable wormholes \cite{Estrada:2023pny}, generalized uncertainty principle \cite{Alencar:2023wyf} and as well as in the D-dimensional extension with higher-curvature corrections \cite{Konoplya:2024kih}. 

Quasinormal modes are an essential characteristic of black holes \cite{Konoplya:2011,Kokkotas:1999}, and these modes are defined by fundamental properties of black holes  rather than external disturbances. They are detectable  through gravitational interferometers \cite{LIGOScientific:2016aoc}. In recent years, the calculation of quasinormal modes (QNMs) for higher-dimensional black holes has captured significant attention for various compelling reasons. For example, the understanding of properties of $D$-dimensional general relativity \cite{Bizon:2005cp, Konoplya:2024hfg} and thermodynamics characteristics \cite{Kunstatter:2002pj}. 

In this study, we investigate the thermodynamic properties and quasinormal modes of Dymnikova black holes within the context of higher dimensions in Einstein's general theory of relativity. We calculate the thermodynamic parameters, including the Hawking temperature and heat capacity, which allowed us to investigate the black hole's stability. Lastly the quasinormal modes with the WKB formula was calculated. The paper is organized as follows. In Sec. II, we present the solution for $D$-dimensional Dymnikova black holes. Sec. III discusses the thermodynamics associated with the black holes. Sec. IV provides a concise overview of the characteristics of the wave equation, and we explore the numerical results obtained for quasinormal frequencies. Finally, we summarize the paper in Sec. V.  We use the positive metric signature $(-, +, +, +)$ and units where universal constants are equated to unity.

\section{A brief review of Dymnikova black holes in higher dimensions}

In this section we briefly review the Dymnikova black hole in higher dimensions following the lines of \cite{Bikash Chandra Paul:2023}, while, at the same time, we introduce a  coordinate transfomation in order to regularize the metric.

The Dymnikova density profile for vacuum in four dimensions can be regarded as the gravitational analog to the Schwinger effect \cite{Dymnikova:1996}. This phenomenon of high-energy quantum electrodynamics occurs when a strong and consistent electric field $(E)$ is applied, leading to vacuum polarization and the creation of particle pairs. The gravitational counterpart of the electron-positron pair production rate in a vacuum arises when we link the electric field to gravitational tension, which is defined by a curvature term given by
\begin{equation}
    \dfrac{E_0}{E} = \dfrac{r^3}{r_{*}}, \qquad \qquad  \qquad \qquad  r_{*}^3 = r_g r_0^2,
\end{equation}
where $E \approx r^{-3}$, while $E_0 = \frac{\pi\hbar m_e^2}{e}$, $r_g$ is the Schwarzschild radius, $r_0$ is connected to the curvature of  the de Sitter core and  $r_{*}$ is a dimensional constant. With these considerations, we have the following density profile:
\begin{equation}
    \rho(r) = \rho_0 \operatorname{exp}\left(-\dfrac{r^3}{r_{*}^3} \right), \quad \quad r_0^2 = \frac{3}{8\pi\rho_0}.
\end{equation}

Paul's generalization to higher dimensions is based on the assumption that the component $T_{0}^{\hspace{0.01cm}0}$ of the momentum energy tensor is given by 
\begin{equation}
    \rho(r) = -T_{0}^{\hspace{0.01cm}0} = \rho_0 e^{-\frac{r^{D - 1}}{r_{*}^{D - 1}}},
\end{equation}
where $T_{\nu}^{\hspace{0.01cm}\mu} = (-\rho, P_r, P_t, \ldots)$, $\rho$ is the energy density, $P_r$ the radial pressure, $P_t$ transverse pressure and $\mu, \nu = 0, 1, \ldots, D - 1$. For black holes, we know that
\begin{equation}
    T_{0}^{\hspace{0.01cm}0} = T_{1}^{\hspace{0.01cm}1}, \qquad T_{\theta_1}^{\hspace{0.01cm}\theta_1} = T_{\theta_2}^{\hspace{0.01cm}\theta_2} = \ldots T_{\theta_{D - 2}}^{\hspace{0.01cm}\theta_{D - 2}},
\end{equation}
when considering the general spherically symmetric metric  with the previous arguments, we obtain the metric potential which is given by
\begin{equation}\label{5}
    f(r) = 1 - \dfrac{r_g^{D - 3}}{r^{D - 3}}\left(1 - e^{-\frac{r^{D - 1}}{r_{*}^{D - 1}}}\right).
\end{equation}
Let $r_{g}^{D - 3} = \frac{2\rho_0}{(D - 1)(D - 2)}r_{*}^{D - 1}$ represent the higher-dimensional Schwarzschild radius in our notation. The generalization of the Dymnikova black hole is obtained as
\begin{equation}\label{6}
    ds^2 = -\left(1 - \frac{R_s(r)}{r^{D - 3}}\right)dt^2 + \dfrac{dr^2}{\left(1 - \frac{R_s(r)}{r^{D - 3}} \right)} + r^2d\Omega_{D - 2}^2,
\end{equation}
where $d\Omega_{D - 2}$ is the line element of a unit $(D-2)$-dimensional sphere \cite{Myers:1986} and $R_S(r)$ is a radius function given by
\begin{equation}
    R_s(r) = r_{g}^{D - 3}\left[1 - \operatorname{exp}\left(-\dfrac{r^{D - 1}}{r_{*}^{D - 1}} \right)\right].
\end{equation}
 The subsequent equations are employed
\begin{equation}
    r_0^2 = \frac{(D - 1)(D - 2)}{16\pi\rho_0} \qquad \text{and} \qquad r_{*}^{D - 1} = r_0^2r_{g}^{D - 3}. 
\end{equation}
This exact spherically symmetric solution of the Einstein field equations  yields the de Sitter solution for $r \ll r_{*}$ and the Schwarzschild solution for $r \gg r_{*}$.  The components of energy momentum tensor are $ T_{0}^{\hspace{0.05cm}0} = T_{1}^{\hspace{0.05cm}1} = -\rho_0 e^{-\frac{r^{D - 1}}{r_{*}^{D - 1}}}$ and
\begin{equation}
    T_{\theta_2}^{\hspace{0.05cm}\theta_2} = \left[\dfrac{D - 1}{D - 2}\left(\dfrac{r}{r_{*}}\right)^{D - 1} - 1 \right]\rho_0 e^{-\frac{r^{D - 1}}{r_{*}^{D - 1}}}.
\end{equation}

The horizons of a black hole are zeros of $g_{tt} = 0$. We have two distinct horizons for $r_g \gg r_0$, the Cauchy horizon $r_{-}$ and the event horizon $r_{+}$, which are located at
\begin{equation}
    r_{-} = r_0\left[1 - O\left(\operatorname{exp}\left(-\frac{r_0}{r_g}\right)\right)\right], \quad r_{+} = r_g\left[1 - O\left(\operatorname{exp}\left(-\frac{r_g^2}{r_0^2} \right) \right) \right].
\end{equation}
They can be eliminated by an appropriate coordinate transformation. In the coordinates connected with the freely falling particles the metric takes the Lemaitre type form
\begin{equation}
    ds^2 = -d\tau^2 + \dfrac{R_s(r)}{r^{D - 3}}dr^2 + r^2d\Omega_{D - 2},
\end{equation}
 $(\lim_{r\rightarrow 0}{(R_s(r)/r^{D - 3})} = 0)$ that is regular both at $r_{+}$ and $r_{-}$ as well as at $r\rightarrow 0$, but is not complete. To find its maximal analytic extension one should introduce the isotropic Eddington-Finkelstein coordinates in which the solution (\ref{6}) is given by
\begin{equation}
    ds^2 = -\left|1 - \dfrac{R_s(r)}{r^{D - 3}}\right|du\hspace{0.1cm}dv + r^2d\Omega_{D - 2}^2.
\end{equation}
So the solution presented here is regular everywhere.

\section{Temperature in Four Dimensions}

In this section, we will initiate the study of the black hole temperature in four dimensions, and we will determine whether there is an analytic condition for the Pure Dymnikova solution to have a remnant, i.e., $T_{+}=0$. This latter is important since, despite getting a  singularity, under the presence of a remnant, the black hole would never completely evaporate. So, in this work we will also explore this problem.

The Dymnikova solution (\ref{5}) in four dimensions is given by
\begin{equation}
    f(r) = 1 - \frac{2M(1-e^{-\frac{r^3}{r_{*}^3}})}{r}.
\end{equation}
To determine the mass function, $M$, we impose $f(r_+)=0$. In order to test the Dymnikova black hole accurately, in this part we will utilize the same setup as the original model \cite{Dymnikova:1992}. In four dimensions, we have  $r_{*}^3=(2Mr_0^2)$. Consequently, we must solve
\begin{equation}
    \frac{r_+}{2M}=1-e^{-\frac{r_+^3}{2Mr_0^2}}
\end{equation}
for $M$. Start by multiplying the above equation by $r_+^2/r_0^2$, so that we obtain
\begin{equation}
    \frac{r_+^3}{2Mr_0^2}=\frac{r_+^2}{r_0^2}-\frac{r_+^2}{r_0^2}e^{-\frac{r_+^3}{2Mr_0^2}}.
\end{equation}
Now we define $y=r_+^3/(2Mr_0^2)$ which leads to
\begin{equation}
    y - \frac{r_+^2}{r_0^2}=-\frac{r_+^2}{r_0^2}e^{-y}.
\end{equation}
Therefore, by multiplying both sides by $e^{y-\frac{r_+^2}{r_0^2}}$, we obtain that
\begin{equation}
    \left(y-\frac{r_+^2}{r_0^2}\right)e^{y-\frac{r_+^2}{r_0^2}} = -\frac{r_+^2}{r_0^2}e^{-\frac{r_+^2}{r_0^2}}.
\end{equation}

The above equation has the shape $ye^y=b$. It can be solved only if $b \geq -1/e$,  with solution given in terms of the the Lambert $W$ function by $y=W_0(b)$. So, we have that
\begin{equation}\label{18}
 y = \frac{r_+^2}{r_0^2} + W_0\left(-\frac{r_+^2}{r_0^2}e^{-\frac{r_+^2}{r_0^2}}\right).
\end{equation}
In other words,
\begin{equation}\label{19}
    \frac{1}{2M} = \frac{1}{r_+} + \frac{r_0^2}{r_+^3}W_0\left(-\frac{r_+^2}{r_0^2}e^{-\frac{r_+^2}{r_0^2}}\right).
\end{equation}
Given the condition $b > -1/e$, we find that the above  solution will occur only if
\begin{equation*}
    \frac{r_+^2}{r_0^2}e^{-\frac{r_+^2}{r_0^2}} < e^{-1}.
\end{equation*}
 
\noindent{Finally, the Hawking Temperature can be properly written as follows,}
\begin{equation}\label{20}
    T_+ = \frac{1}{4\pi r_0}\left[\dfrac{r_0}{r_+} - \dfrac{3r_+}{r_0}\left(1 - \dfrac{r_+}{r_g}\right)\right].
\end{equation}

The  “remnant possibility” refers to the scenario where the Hawking temperature vanishes,  indicating a remnant black hole that does not evaporate completely. The Black hole remnant could serve as potential dark matter candidates, as their stability and non-evaporating nature make them viable for such a role \cite{Chen:2002tu, Dymnikova:2010zz, Adler:2001vs}. Mathematically, this occurs when the horizon radius satisfies the condition for zero Hawking temperature, $T_+=0$, this is determined by
\begin{equation}
    W_0\left(-\frac{r_+^2}{r_0^2}e^{-\frac{r_+^2}{r_0^2}}\right) = -\dfrac{1}{3},
\end{equation}
inserting this into Eq.(\ref{19}), we get the remnant mass
\begin{equation}
     M = \dfrac{3r_+^3}{2(3r_+^2 - r_0^2)}.
\end{equation}
Next we will perform the same analysis for the Dymnikova solution in $D$ dimensions.

\section{Thermodynamics of Higher Dimensional Dymnikova black hole}

Now, we calculate the thermodynamics quantities associated with the D-dimensional Dymnikova black holes (\ref{6}). The mass of the black hole can be determined based on the horizon radius $r_+$ by solving the equation $f(r_+) = 0$, resulting in
\begin{equation}\label{13}
    M = \dfrac{(D-2)\Omega_{D - 2}r_+^{D - 3}}{16\pi\left[1 + \frac{r_0^2}{r_+^2}W_0\left(-\dfrac{r_+^2}{r_0^2}e^{-r_+^2/r_0^2}\right)\right]}.
\end{equation} 
where
\begin{equation}
    \Omega_{D - 2} = \dfrac{2\pi^{\frac{D-1}{2}}}{\Gamma\left(\frac{D-1}{2}\right)}, \qquad \qquad r_g^{D - 3} = \dfrac{16\pi M}{(D - 2)\Omega_{D - 2}}.
\end{equation}
The black hole possesses a Hawking temperature, which can be derived from the surface gravity provided by
\begin{equation}
    \kappa = \left.\dfrac{1}{2}\dfrac{df(r)}{dr}\right|_{r = r_{+}}.
\end{equation}
Subsequently, Inserting the value of mass from Eq.(\ref{13}) in the surface gravity, we have that the Hawking temperature $(T_{+} = \kappa/2\pi)$ for Dymnikova black holes is expressed as follows:
\begin{equation}
    T_{+}  = \dfrac{1}{4\pi r_0}\left[\dfrac{(D - 3)r_0}{r_+} - \dfrac{(D - 1)r_+}{r_0}\left(1 - \dfrac{r_+^{D - 3}}{r_g^{D - 3}}\right) \right].
\end{equation}
\begin{figure*}[ht]
    \centering
    \begin{subfigure}{0.5\textwidth}
    \includegraphics[scale=0.9]{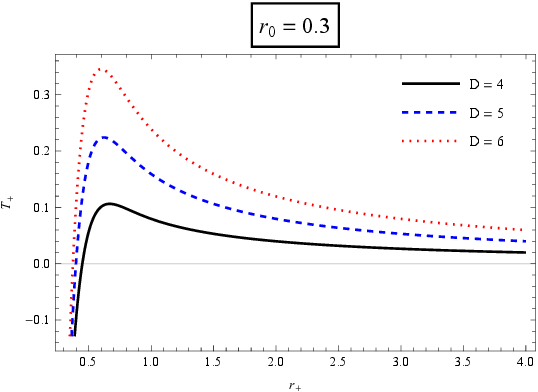}
  \end{subfigure}%
  \begin{subfigure}{0.5\textwidth}
    \includegraphics[scale=0.9]{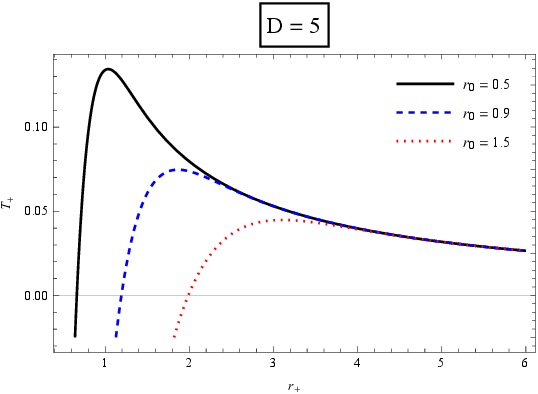}
  \end{subfigure}
  \caption{Plotting the Hawking temperature ($T_+$) as a function of the event horizon radius ($r_+$).}
  \label{Figure 1}
\end{figure*}
When $r_{+} \gg r_{0}$ we recover the spherically-symmetric $D$ dimensional black hole Hawking temperature \cite{HarrisandKanti:2003}. For $D = 4$, this reduces to the  Hawking temperature for Dymnikova found in (\ref{20}). We plot in Fig.\ref{Figure 1} the behaviour of the Hawking temperature for the D-dimensional Dymnikova black hole for various dimensions. In the left panel we have set $r_{0}=0.3$ and considered three values for $D$, namely, $D=4$, $D=5$ and $D=6$. The behavior of the Hawking temperature for the three considered dimensions is similar. All cases exhibit a phase transition of zeroth order in which the temperature vanishes and the black hole evaporation halts at finite horizon radii. Therefore, remnant masses appear as a result of such phase transitions. We can see in Fig.\ref{Figure 1} (left panel) that by increasing the dimension, the value of $r_{+}$ at which the Hawking temperature vanishes becomes smaller. Hence, the remnant mass is affected by the dimension. In the right panel of Fig.\ref{Figure 1} we have set $D=5$ and considered three values of $r_{0}$, namely, $r_{0}=0.3$, $r_{0}=0.9$ and $r_{0}=1.5$. We can see that the remnant masses are also affected by the increasing value of $r_{*}$ for a given dimension, so that as we increase the value of $r_{0}$, the value of $r_{+}$ at which the Hawking temperature vanishes becomes greater.

\begin{figure}[ht]
    \centering
    \includegraphics[scale=0.9]{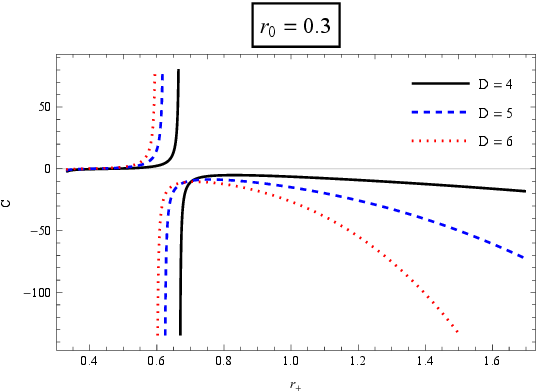}
    \caption{Plotting the heat capacity ($C$) as a function of the event horizon radius ($r_+$) for the Dymnikova black hole across different dimensions and with fixed $r_*$.}
    \label{Figure 2}
\end{figure}

The determination of potential phase transitions in the black hole hinges on the criterion for a change in heat capacity sign. A positive heat capacity ($C > 0$) is a signature of local stability against thermal fluctuations, whereas a negative heat capacity ($C < 0$) indicates local instability. The expression for heat capacity is as follows:
\onecolumngrid
\begin{equation}\label{28}
    C  = \frac{\left(\frac{D - 2}{4}\right)\Omega_{D - 2} r_{+}^{D - 2}\left[(D - 3) + \frac{2W_0\left(-\frac{r_{+}^2}{r_0^2}e^{-r_{+}^2/r_0^2}\right)}{1 + W_0\left(-\frac{r_{+}^2}{r_0^2}e^{-r_{+}^2/r_0^2}\right)}\right]}{\left[1 + \frac{r_0^2}{r_{+}^2}W_0\left(-\frac{r_{+}^2}{r_0^2}e^{-r_{+}^2/r_0^2}\right)\right]\left\{-(D - 3) - (D - 1)W_0\left(-\frac{r_{+}^2}{r_0^2}e^{-r_{+}^2/r_0^2}\right)\left[1 - \frac{2\left(1 - \frac{r_{+}^2}{r_0^2}\right)}{1 + W_0\left(-\frac{r_{+}^2}{r_0^2}e^{-r_{+}^2/r_0^2}\right) } \right]  \right\} }.
\end{equation}
\twocolumngrid
Figure \ref{Figure 2} displays plots of the heat capacity (\ref{28}) for a fixed value of $r_{0}$ across different dimensions. The heat capacity has one Davies point \cite{Davies}, being such point related to the maximum of the Hawking temperature. Hence, the Davies point set the value of $r_{+}$ in which the D-dimensional Dymnikova black hole exhibit a phase transition, and as we can see, the Davies point is dimensional-dependent. As we increase the value of the dimension, the position of the Davies point is shifted to the left, so that the phase transition occur for smaller values of $r_{+}$.

\section{Quasinormal Modes}
\begin{figure*}[t]
    \centering
    \begin{subfigure}{0.5\textwidth}
    \includegraphics[scale=0.9]{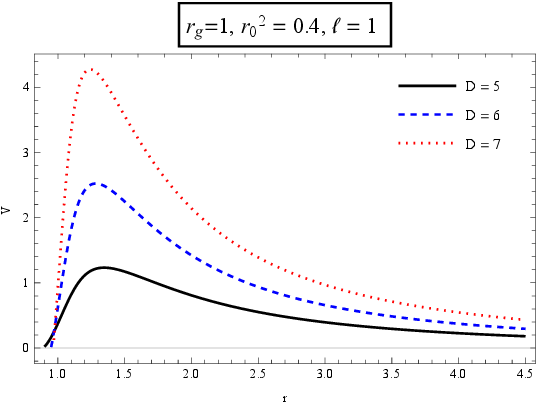}
  \end{subfigure}%
  \begin{subfigure}{0.5\textwidth}
    \includegraphics[scale=0.9]{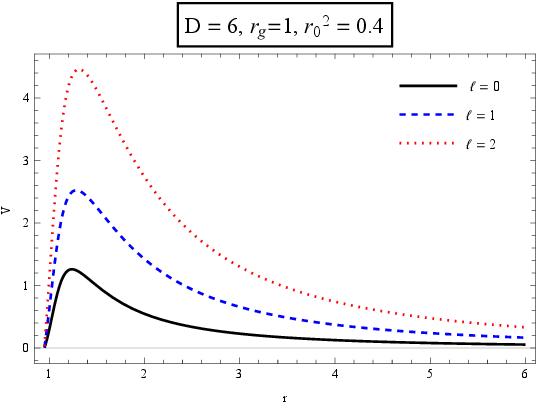}
  \end{subfigure}
  \begin{subfigure}{0.5\textwidth}
    \includegraphics[scale=0.9]{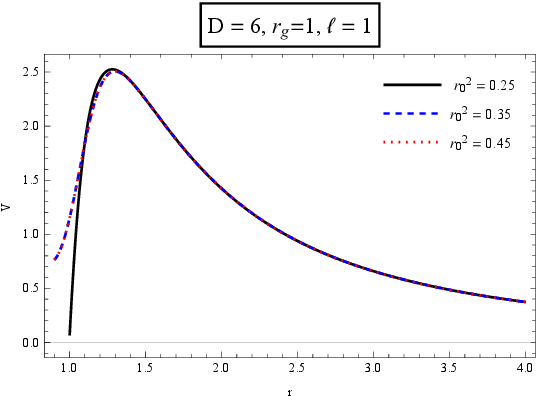}
  \end{subfigure}%
  \begin{subfigure}{0.5\textwidth}
    \includegraphics[scale=0.9]{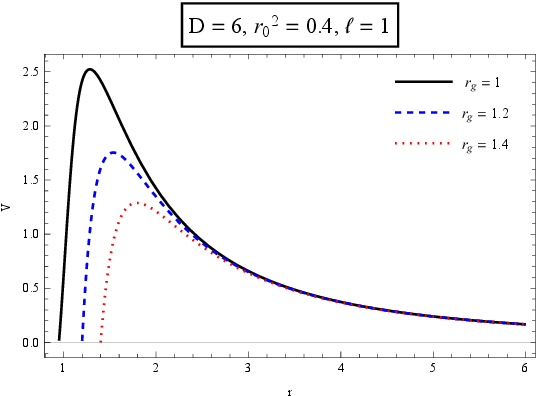}
  \end{subfigure}
  \caption{Effective potential barrier for scalar perturbations vs. radial coordinate for the parameters shown in the panels.}
  \label{Figure 3}
\end{figure*}

Initially, we will provide a concise overview of the essential components required to calculate the quasinormal modes for scalar perturbations. Let's assume the existence of a real, electrically neutral, massive scalar field, $\Phi$, that is canonically coupled to gravity. Now, consider its propagation in a fixed gravitational background. The Klein-Gordon equation is expressed as follows
\begin{equation}\label{29}
    \dfrac{1}{\sqrt{-g}}\partial_{\mu}(-\sqrt{-g}\hspace{0.1cm}g^{\mu\nu}\partial_{\nu})\Phi = m^2\Phi.
\end{equation}
where $m$ is the mass of the scalar field. We employ the separation of variables and introduce the following supposition to solve the the previous equation
\begin{equation}
    \Phi(t, r, \theta_1, \ldots, \theta_{D - 2}) = e^{-i\omega t}\dfrac{\psi(r)}{r^{(D - 2)/2}}\Tilde{Y}_{l}(\Omega),
\end{equation}
being $\omega$ the frequency and the generalization of the spherical harmonics is given by $\Tilde{Y}_{l}$ \cite{Higuchi:1986}. The equations in (\ref{29}) take the
Schrödinger wavelike form \cite{Panotopoulos:2020}:
\begin{equation}
    \dfrac{d^2\psi}{dx^2} + U(x,\omega)\psi = 0,
\end{equation}
 where $U(x, \omega)$ is the effective potential that depends on the frequency of the wave and the "tortoise coordinate" denoted as $x$, which is defined in the following manner:
\begin{equation}
    dx \equiv \dfrac{dr}{1 - \frac{R_s(r)}{r^{D - 3}}}.
\end{equation}
As $x$ approaches negative infinity, it signifies the presence of the event horizon, while the limit as $x$ approaches positive infinity corresponds to spatial infinity. By definition, the wave is ingoing when
\begin{equation}
    \psi_{in}(x\rightarrow\pm\infty) \propto 
    \left\{\begin{matrix}
       & e^{-ik_{\pm}z}, && \omega > 0; \\ 
        & e^{ik_{\pm}x}, && \omega < 0;
    \end{matrix}\right.
\end{equation}
and outgoing for
\begin{equation}
    \psi_{out}(x\rightarrow\pm\infty) \propto 
    \left\{\begin{matrix}
       & e^{ik_{\pm}z}, && \omega > 0; \\ 
        & e^{-ik_{\pm}x}, && \omega < 0
    \end{matrix}\right.
\end{equation}
The wave number $k_{\pm}(\omega)$, which is greater than zero, satisfies the dispersion relations.  Typically, the effective potential takes the form
\begin{equation}
    U(x, \omega) = V(x) - \omega^2.
\end{equation}
The potential $V(x)$ for the Klein-Gordon equation have the form\cite{Harmark-Schiappa:2010}
\begin{equation}
    \begin{aligned}
    V(r) = \left(1 - \dfrac{R_s(r)}{r^{D - 3}}\right)& \left[m^2 + \dfrac{\ell(\ell + D - 3)}{r^2} - \dfrac{D - 2}{2r}\dfrac{d}{dr}\left(\dfrac{R_s(r)}{r^{D - 3}}\right) + \right. \\& \left. \quad \dfrac{(D - 2)(D - 4)}{4r^2}\left(1 - \dfrac{R_s(r)}{r^{D - 3}}\right) \right],
\end{aligned}
\end{equation}
where $\ell \geq 0$ is the orbital angular momentum. The potential barrier for different dimensions is shown in Figure \ref{Figure 3}, for fixed $\{r_g, r_{0}^2, \ell\}$ and different values of dimensions $D$ (top-left graph), it is evident that with an increase in $D$, the potential's maximum also increases. For fixed $\{D, r_g, r_{0}^2\}$ and varying the orbital angular momentum $\ell$ (top-right graph),  it is evident that with an increase in $\ell$, the potential's maximum also increases, moving consistently to the right. For fixed $\{D, r_g, \ell\}$ and varying $r_{0}^2$ (bottom-left graph), it is evident that with an increase in $r_{0}^2$, the potential's maximum remains almost the same, with only a slight modification. Lastly, for fixed $\{D, r_{*}, \ell\}$ and varying the Schwarzschild radius (bottom-right graph), it is evident that with an increase in $r_g$, the potential's maximum decreases, moving consistently to the right.

When purely outgoing and ingoing waves are enforced at both spatial infinities in the fundamental wave equation, it gives rise to the quasinormal modes denoted as $\omega$ in asymptotically flat black holes. The WKB formula provides a closed form for the quasinormal modes as,
\begin{equation}
    \begin{aligned}
    \omega^2 & = V_0 + A_2(\mathcal{K}^2) + A_4(\mathcal{K}^2) + A_6(\mathcal{K}^2) + \ldots \\&
    - i\mathcal{K}\sqrt{-2V_0''}(1 + A_3(\mathcal{K}^2) + A_5(\mathcal{K}^2) + A_7(\mathcal{K}^2) + \ldots),
\end{aligned}
\end{equation}
where $A_k(\mathcal{K}^2)$ are polynomials of the derivatives $U'',U''', \ldots$ that can be found in \cite{Konoplya:2019hlu} and $\mathcal{K}$ is for QN modes equal to
\begin{equation*}
    \mathcal{K} = 
    \left\{\begin{matrix}
        & +n + \frac{1}{2}, \quad  && \text{$\operatorname{Re}(\omega) > 0$;}  \\ \\
        & -n - \frac{1}{2}, \hspace{0.5cm}   && \hspace{0.2cm} \text{$\operatorname{Re}(\omega) < 0$;}
    \end{matrix}\right.
\end{equation*}
with $n = 0, 1, 2, 3,\ldots$, and $V_0, V_0'', V_0''', \ldots$ are, respectively, the value and higher derivatives of the potential $V(x)$ in the maximum.

Increasing the WKB order does not necessarily result in a more accurate approximation of the frequency. Typically, the error of the WKB formula approximation is assessed by comparing two consecutive orders. To estimate the error for $\omega_k$ obtained through the WKB formula of order $k$, we employ the quantity
\begin{equation}\label{38}
    \Delta_k = \dfrac{|\omega_{k + 1} - \omega_{k - 1}|}{2}.
\end{equation}

 In order to increase the accuracy of the WKB formula, we will follow the procedure of Matyjasek and Opala \cite{Matyjasek:2017psv, Mathematica} and use the Padé approximants. 
 

Precise analytical formulations for the quasinormal spectra of black holes are attainable only in select cases, such as when the differential equation corresponding to the radial component of the wave function can be transformed into the Gauss hypergeometric function and in instances for potentials with the Pöschl-Teller shape. For Dymnikova black hole, due to the non-trivial characteristics of equation (\ref{29}), it becomes imperative to resort to numerical methods to calculate the corresponding QN modes. Hence, several methods have emerged for this goal, in this paper we'll apply the well-established WKB, which is commonly used for computing
the quasinormal modes of black holes,  approach to determine the QN frequencies. 

\begin{figure*}[ht]
    \centering
    \begin{subfigure}{0.5\textwidth}
    \includegraphics[scale=0.9]{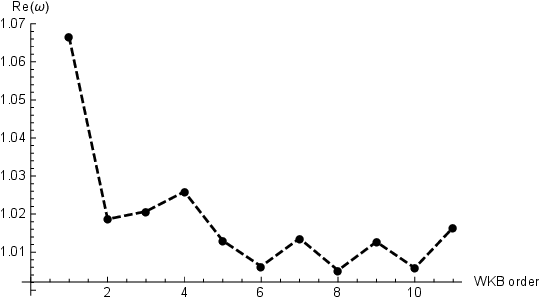}
  \end{subfigure}%
  \begin{subfigure}{0.5\textwidth}
    \includegraphics[scale=0.9]{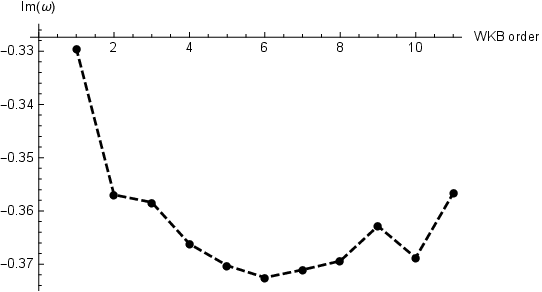}
  \end{subfigure}
  \begin{subfigure}{0.5\textwidth}
    \includegraphics[scale=0.9]{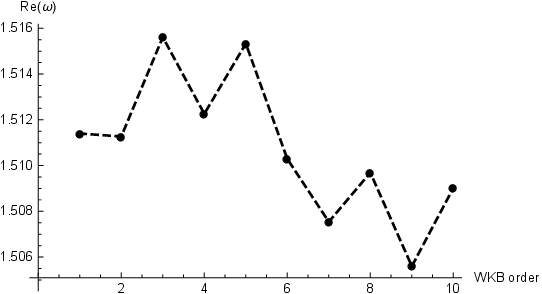}
  \end{subfigure}%
  \begin{subfigure}{0.5\textwidth}
    \includegraphics[scale=0.9]{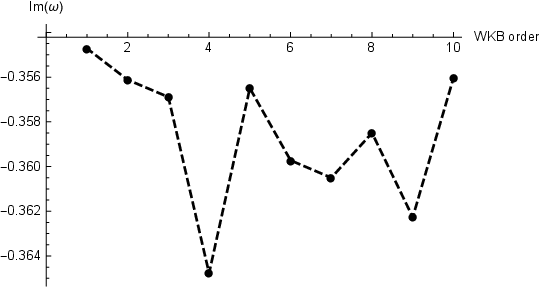}
  \end{subfigure}
  \caption{$\omega_{Re}$ (left) and $\omega_{Im}$ (right) as a function of WKB order of the formula with which it was obtained for $n = 0, D = 5, r_g = 1, r_{0}^2 = 0.2$, $\ell = 1$ (top) and $\ell = 2$ (bottom)  modes for massless scalar field.}
  \label{Figure 4}
\end{figure*}

Typically, the WKB formula tends to be more accurate when $\ell$ is higher and both $n$ and $D$ are lower. Furthermore, for a specified orbital angular momentum $\ell$, only values of $n \leq \ell$ will be taken into consideration.  In Fig.\ref{Figure 4}  we can see that in order to find the fundamental mode with sufficiently high precision, the highest level of precision is attained by employing high WKB orders in conjunction with the Padé approximation.

In table I we summarize accuracy of the WKB formula for $\ell = 1$ in $4, 5$ and $6$ dimensions.  We see that the error can be very well estimated by comparing frequencies, obtained with the help of the WKB formula at different orders. The quantity $\Delta_k$, introduced in (\ref{38}), not only allows one to find the order of the absolute error but also to determine the order, which gives the most accurate approximation for the quasinormal modes. Our findings indicate that, within the range of parameters considered, the black hole remains stable against scalar perturbations based on the QNMs computations, this holds true due to negative $\operatorname{Im}(\omega)$.

\begin{table}[ht!]
    \centering
    \begin{tabular}
    {|c|c|c|}\hline
        $r_0^2$ & $\omega$ & $\text{Sixth order WKB}$  \\ \hline
        \multicolumn{3}{|c|}{ $\ell = 1$, $n = 0$, $D = 4$} \\ \hline
        0.20 &  0.583422 - 0.195365i  & 0.579949 - 0.201839i \\ \hline 
        0.25 &  0.580694 - 0.193959i & 0.574068 - 0.190187i \\ \hline 
        0.30 &  0.575456 - 0.194747i & 0.571264 - 0.195078i \\ \hline 
        0.35 &  0.566952 - 0.191897i & 0.555443 - 0.200196i \\ \hline 
        0.40 &  0.562671 - 0.179931i & 0.547077 - 0.184825i \\ \hline 
        0.45 &  0.547112 - 0.176065i & 0.555613 - 0.180324i \\ \hline 
        \multicolumn{3}{|c|}{ $\ell = 1$, $n = 0$, $D = 5$} \\ \hline
        0.20 &  1.01299 - 0.370259i & 1.00626 - 0.372503i \\ \hline 
        0.25 &  1.00921 - 0.366918i & 1.01385 - 0.367913i \\ \hline 
        0.30 &  1.00237 - 0.358466i & 1.00814 - 0.368051i \\ \hline 
        0.35 &  1.00033 - 0.350568i & 1.00719 - 0.363288i \\ \hline 
        0.40 &  0.981471 - 0.356504i & 0.980692 - 0.366012i \\ \hline 
        0.45 &  0.982178 - 0.33442i & 0.964823 - 0.362666i \\ \hline 
        \multicolumn{3}{|c|}{ $\ell = 1$, $n = 0$, $D = 6$} \\ \hline
        0.20 &  1.44322 - 0.517553i & 1.44063 - 0.525754i \\ \hline 
        0.25 &  1.43482 - 0.505017i & 1.4427 - 0.52096i \\ \hline 
        0.30 &  1.42936 - 0.510123i & 1.44053 - 0.520656i \\ \hline 
        0.35 & 1.41782 - 0.480749i & 1.42884 - 0.5221i \\ \hline 
        0.40 &  1.4067 - 0.50002i & 1.42285 - 0.520919i \\ \hline 
        0.45 &  1.39392 - 0.48167i & 1.38135 - 0.519523i  \\ \hline 
    \end{tabular}
    \caption{QN modes of the massless scalar field for different dimensions and $r_0^2$, $\ell = 1$ and $r_g = 1$, computed using the WKB formula.}
    \label{Table I}
\end{table}

\section{Final Remarks}

Exploring general relativity in higher dimensions promises valuable insights into the theory's nature, particularly in the context of black holes. In this work, we present an exact static spherically symmetric Dymnikova black hole that is regular within an arbitrary D-dimensional spacetime. We analyzed the solution, identifying potential horizons. Additionally, we verified the spacetime's regular structure and conducted a thermodynamic analysis. We investigate the behaviour of the Hawking temperature by varying the dimensions. All dimensions exhibit a phase transition of zeroth order in which the temperature vanishes and the black hole evaporation halts at finite horizon radii. Therefore, remnant masses appear as a result of such phase transitions. We could see that by increasing the dimension, the value of $r_{+}$ at which the Hawking temperature vanishes becomes smaller. Hence, the remnant mass is affected by the dimension. We have also studied the heat capacity, since the determination of potential phase transitions in the black hole hinges on the criterion for a change in heat capacity sign. A positive heat capacity ($C > 0$) is a signature of local stability against thermal fluctuations, whereas a negative heat capacity ($C < 0$) indicates local instability. We have shown that the heat capacity has one Davies point, being such point related to the maximum of the Hawking temperature. Hence, the Davies point set the value of $r_{+}$ in which the D-dimensional Dymnikova black hole exhibit a phase transition, and we have also shown that the Davies point is dimensional-dependent. As we increase the value of the dimension, the position of the Davies point is shifted to the left, so that the phase transition occur for smaller values of $r_{+}$.

The study of quasi-normal modes of a Dymnikova black hole offer a unique window into its fundamental properties and gravitational behavior. In this paper, we have illustrated graphical representations of the effective potential barrier with respect to the radial coordinate, $r$, by individually varying the set of parameters ${D, r_g, r_{*}, \ell}$. Following this, we numerically computed the QNMs using the WKB method. Our findings indicate that, within the range of parameters considered, the black hole remains stable against scalar perturbations based on the QNMs computations, this holds true due to negative $\operatorname{Im}(\omega)$. The results presented in this work expand upon the Dymnikova Black hole, implying potential relevance within the framework of string theory.

\acknowledgments

\hspace{0.5cm} The authors thank the Coordena\c{c}\~{a}o de Aperfei\c{c}oamento de Pessoal de N\'{i}vel Superior (CAPES). JF would like to thank the Fundação Cearense de Apoio ao Desenvolvimento Cient\'{i}fico e Tecnol\'{o}gico (FUNCAP) under the grant PRONEM PNE0112-00085.01.00/16 for financial support and Gazi University for the kind hospitality.


\begin{thebibliography}{99}



\bibitem{Finkelstein:1958}
Bekenstein, J.D.,
   Phys. Rev. \textbf{110}, 965 (1958)

\bibitem{EventHorizonTelescope:2019dse}
K.~Akiyama \textit{et al.} [Event Horizon Telescope],
Astrophys. J. Lett. \textbf{875}, L1 (2019)


\bibitem{Bekenstein:1972}
Bekenstein, J.D.,
   Lett. Nuovo Cim. \textbf{4}, 737 (1972)


\bibitem{Bekenstein:1973}
Bekenstein, J.D.,
  Phys. Rev. D \textbf{7}, 2333 (1973)

\bibitem{Bekenstein:1974}
Bekenstein, J.D.,
  Phys. Rev. D \textbf{10}, 3292 (1974)


\bibitem{Chen:2002tu}
Chen, Pisin and Adler, Ronald J.,
Nucl. Phys. B Proc. Suppl. \textbf{124}, 103-106 (2003).


\bibitem{Dymnikova:2010zz}
Dymnikova, Irina and Korpusik, Michal,
Phys. Lett. B \textbf{685}, 12-18 (2010).


\bibitem{Adler:2001vs}
Adler, Ronald J. and Chen, Pisin and Santiago, David I.,
Gen. Rel. Grav. \textbf{33}, 2101--2108 (2001).


\bibitem{Hawking:1975}
Hawking, S.W.,
  Commun. Math. Phys. \textbf{43}, 199 (1975)



\bibitem{Sakharov:1966}
A. D. Sakharov,
   Sov. Phys. JETP \textbf{22}, 241 (1966)

\bibitem{Gliner:1966}
E. B. Gliner,
   Sov. Phys. JETP \textbf{22}, 378 (1966)

\bibitem{bardeen}
J. M. Bardeen,
Proceedings of the International Conference GR5, Tbilisi, USSR, p. 174. Tbilisi University Press, (1968)

\bibitem{Christiansen:2022ebo}
H.~R.~Christiansen, M.~Estrada, M.~S.~Cunha, J.~Furtado and C.~R.~Muniz,
Int. J. Mod. Phys. D \textbf{32}, no.07, 2350041 (2023)

\bibitem{Furtado:2022tnb}
J.~Furtado and G.~Alencar,
Universe \textbf{8}, no.12, 625 (2022)


\bibitem{Rodrigues:2015}
Rodrigues, ~Manuel ~E. and Junior, ~Ednaldo ~L. ~B. and Marques, ~Glauber ~T. and Zanchin, ~Vilson ~T.
Phys. Rev. D \textbf{94}, no.02, 024062 (2016)


\bibitem{Dymnikova:1992}
Dymnikova, I.,
   Gen. Rel. Gravity \textbf{24}, 235 (1992)

\bibitem{Lan:2023cvz}
Lan, Chen and Yang, Hao and Guo, Yang and Miao, Yan-Gang,
   Int. J. Theor. Phys. \textbf{62}, 202 (2023)


\bibitem{Bikash Chandra Paul:2023}
B. C. Paul,
Eur. Phys. J. Plus \textbf{138}, 633 (2023).

\bibitem{Nashed:2003}
G. G. L. Nashed,
    Chaos Solitons Fractals \textbf{15}, 841 (2003).


\bibitem{Dymnikova:2005}
Dymnikova, I. and E. Galaktionov,
    Class. Quant. Grav. \textbf{22}, 2331 (2005).


\bibitem{Konoplya:2023}
R. A. Konoplya, Z. Stuchlík, A. Zhidenko, and A. F. Zinhailo
    Phys. Rev. D \textbf{107}, 104050 (2023).



\bibitem{Sharif:2022}
Sharif, M. and Khan, Amjad,
Mod. Phys. Lett. A \textbf{37}, 2250049 (2022).

\bibitem{Estrada:2023pny}
M.~Estrada and C.~R.~Muniz,
JCAP \textbf{03}, 055 (2023)

\bibitem{Alencar:2023wyf}
G.~Alencar, M.~Estrada, C.~R.~Muniz and G.~J.~Olmo,
JCAP \textbf{11}, 100 (2023)

\bibitem{Konoplya:2011}
Konoplya, R. A. and Zhidenko, A.,
Rev. Mod. Phys.  \textbf{83}, 793-836 (2011).

\bibitem{Konoplya:2024kih}
Konoplya, R. A. and Zhidenko, A.,
Phys. Lett. B \textbf{856}, 138945 (2024).
     

\bibitem{Kokkotas:1999}
Kostas D. Kokkotas and Bernd G. Schmidt.,
Living Rev. Rel.  \textbf{2}, 2 (1999).

\bibitem{LIGOScientific:2016aoc}
Abbott, B. P. and others,
Phys. Rev. Lett.  \textbf{116} (6):061102 (2016).

\bibitem{Bizon:2005cp}
Bizon, Piotr and Chmaj, Tadeusz and Schmidt, Bernd G.,
Phys. Rev. Lett.  \textbf{95}, 071102 (2005).


\bibitem{Konoplya:2024hfg}
R.~A.~Konoplya and A.~Zhidenko,
Phys. Rev. D \textbf{109}, 104005 (2024). 


\bibitem{Kunstatter:2002pj}
Kunstatter, Gabor,
Phys. Rev. Lett.  \textbf{90}, 161301 (2003).



\bibitem{Dymnikova:1996}
Dymnikova, I.,
 Int.J.Mod.Phys.D \textbf{6}, 529 (1996).





\bibitem{Myers:1986}
Myers, Robert C. and Perry, M. J.,
Annals Phys. \textbf{172}, 304 (1986).





\bibitem{HarrisandKanti:2003}
C. M. Harris and P. Kanti,
JHEP \textbf{0310}, 014 (2003).


\bibitem{Davies}
P. C. W. Davies,
Rep. Prog. Phys. {\bf 41} 1313 (1978).



\bibitem{Higuchi:1986}
Higuchi, Atsushi,
J. Math. Phys. \textbf{28}, 1553 (1987).  




\bibitem{Panotopoulos:2020}
G. Panotopoulos,
  Axioms, \textbf{10}, 33 (2020).


\bibitem{Harmark-Schiappa:2010}
Harmark, T. and Natario, J. and Schiappa, R.,
  Adv. Theor. Math. Phys., \textbf{14}, 727 (2010).

\bibitem{Konoplya:2019hlu}
Konoplya, R. A. and Zhidenko, A. and Zinhailo, A. F.,
  Class. Quant. Grav., \textbf{36}, 155002 (2019).


\bibitem{PhysRevD.35.3621}
Iyer, Sai and Will, Clifford M.,
 Phys. Rev. D, \textbf{35}, 12 (1987).

\bibitem{Matyjasek:2017psv}
Matyjasek, Jerzy and Opala, Micha\l{},
 Phys. Rev. D, \textbf{96}, 024011 (2017).

\bibitem{Mathematica}
The Mathematica® package containing the WKB formula is accessible for download at https://goo.gl/nykYGL. 



\end{thebibliography}
\end{document}